# Phonon modes of monoclinic BiB$_3$O$_6$

**A. Gössling[*,1], T. Möller[1], W.-D. Stein[1], P. Becker[3], L. Bohatý[3], and M. Grüninger[1,2]**

[1] 2nd Physical Institute, University of Cologne, Zülpicher Str. 77, D-50937 Cologne, Germany
[2] 2nd Physical Institute, RWTH Aachen, Huyskensweg, D-52056 Aachen, Germany
[3] Institute of Crystallography, University of Cologne, Zülpicher Str. 49b, D-50674 Cologne, Germany

We present a detailed study of the phonon modes of the monoclinic compound BiB$_3$O$_6$ based on polarized reflectivity measurements on single crystals. The spectra are analysed by means of a generalized Drude-Lorentz model, which allows us to resolve the modes of $A$ and $B$ symmetry.

**1 Introduction**   Bismuth triborate, BiB$_3$O$_6$, a polar, non-ferroelectric crystal has attracted interest in the last years due to its outstanding nonlinear optical properties [1-3]. The large optical nonlinearities open up a rich field of applications for frequency conversion of laser light via $\chi^{(2)}$- and $\chi^{(3)}$-processes, e.g. phase-matched second harmonic-generation (SHG) or optical parametric oscillation (OPO) and stimulated Raman scattering (SRS).

The exceptional optical non-linearities of BiB$_3$O$_6$ have been attributed to the bonds of the [BO$_3$] units and to a lone-pair electron at the Bi ion [4]. Detailed studies of the lattice dynamics are required for a quantitative description of these bonds. Infrared (IR) and Raman studies of the phonons are so far only available at room temperature [3,5-8]. In particular, there is no polarization-dependent infrared study, but a polarization analysis is essential for an accurate determination of the phonon frequencies in monoclinic crystals [9]. In this letter we present a detailed investigation of the linear optical response of BiB$_3$O$_6$ in the phonon range for different polarizations at $T = 20$ K and 300 K. The data are analyzed in terms of a generalized Drude-Lorentz model, which allows us to obtain the frequency, the damping, the strength and the orientation of the dipole moment of each phonon mode.

**2 Experimental**   Milky-colored, right-handed single crystals of BiB$_3$O$_6$ with dimensions of 15·15·4 mm$^3$ and 8·8·2 mm$^3$ were grown using the top seeded growth technique [8,10]. The crystal structure with space group symmetry $C2$ ($C_2^3$) ( $a$ = 7.116(2) Å, $b$ = 4.993(2) Å, $c$ = 6.508(3) Å, $\beta$ = 105.62(3)° ) consists of sheets of corner-sharing [BO$_3$]- and [BO$_4$]-units in a ratio of 2:1. These sheets are separated by sheets of six-fold coordinated Bi [11]. We performed reflectivity measurements at $T = 20$ K and 300 K in the spectral range of 50-8000 cm$^{-1}$ at quasi-normal incidence. Using a Bruker IFS 66v/S Fourier-transform spectrometer the spectra were measured for different polarization angles on polished (010) and (100) surfaces of BiB$_3$O$_6$. As a reference we used an Au mirror. The variation of the polarization angle was realized by rotating the polarizer.

**3 Phonon modes**   The factor group analysis [12] yields the following irreducible representations for the space group $C2$:

$$\Gamma_{BiB_3O_6} = \Gamma_{C_2^3} = 14A + 16B \qquad (1)$$

After subtraction of the three acoustic modes $A+2B$, our group theoretical analysis predicts 13$A$ and 14$B$ optical phonon modes. Due to the lack of a center of inversion, $A$ and $B$ modes are active both in Raman [6,7] and in IR spectroscopy. The linear dielectric response (without external magnetic field) of a monoclinic sample can be described by the tensor

$$\hat{\varepsilon} = \begin{pmatrix} \varepsilon_{xx} & 0 & \varepsilon_{xz} \\ 0 & \varepsilon_{yy} & 0 \\ \varepsilon_{xz} & 0 & \varepsilon_{zz} \end{pmatrix} \qquad (2).$$

Note, that we assume $y \parallel b$, $z \parallel c$, and $x \perp c$ lying in the $ac$ plane, where $a$, $b$, and $c$ are the crystallographic axes (see inset of Fig. 1(a)). In order to determine the phonon modes we follow the procedure described in Ref. [9]. The tensor (2) can be decomposed into a scalar $\varepsilon_b$ along the $b$ axis and a two-dimensional tensor $\hat{\varepsilon}_{ac}$ within the $ac$ plane. Since the $b$ axis is perpendicular to the $ac$ plane, the $A$-symmetry modes can be probed by measuring with the incident electric field $E$ parallel to the $b$ axis (e.g. on a (100) surface). Determining the $B$-symmetry modes requires the analysis of at least three polarization directions (with $E \parallel ac$ plane on a (010) surface) because the angle between the $a$ and the $c$ axis deviates from 90°. Reflectivity spectra at $T = 300$ K are shown for $E \parallel ac$ in Fig. 1(a-c) and for $E \parallel b$ in Fig. 1(d). The phonons extend up to almost 1500 cm$^{-1}$ due to the small mass of the B ions.

The dispersion of the scalar $\varepsilon_b$ is described by the a sum of oscillators (Drude-Lorentz model), whereas the tensor $\hat{\varepsilon}_{ac}$ is described by a generalized Drude-Lorentz model:

$$\varepsilon_b(\omega) = \varepsilon_{yy} = \varepsilon_b^\infty + \sum_{i,A} \frac{\omega^2_{p,i}}{\omega^2_{0,i} - \omega^2 - i\gamma_i\omega}$$

$$\hat{\varepsilon}_{ac}(\omega) = \begin{pmatrix} \varepsilon_{xx} & \varepsilon_{xz} \\ \varepsilon_{xz} & \varepsilon_{zz} \end{pmatrix} = \hat{\varepsilon}_{ac}^\infty + \sum_{i,B} \frac{\omega^2_{p,i}}{\omega^2_{0,i} - \omega^2 - i\gamma_i\omega} \times \begin{pmatrix} \cos^2\theta_i & \sin\theta_i\cos\theta_i \\ \sin\theta_i\cos\theta_i & \sin^2\theta_i \end{pmatrix} \quad (3)$$

Here, $\varepsilon_b^\infty$ and $\hat{\varepsilon}_{ac}^\infty$ are the high-frequency dielectric constants, $\omega_{0,i}$ is the transverse frequency, $\omega_{p,i}$ the plasma frequency, $\gamma_i$ the damping of the $i$-th oscillator, and $\theta_i$ (for the $B$ modes) the angle between the dipole moment and the $x$ axis. Due to the low symmetry, the orientation of the principal axes of $\hat{\varepsilon}_{ac}$ depends on $\omega$ and is different for Re$\{\hat{\varepsilon}_{ac}\}$ and Im$\{\hat{\varepsilon}_{ac}\}$. The rotation angles $\phi_{Re}$ and $\phi_{Im}$ can be calculated by diagonalizing Re$\{\hat{\varepsilon}_{ac}\}$ and Im$\{\hat{\varepsilon}_{ac}\}$ by two different rotation matrices. An example for the relationship between $\phi_{Im}$, $\phi_{Re}$, and $\theta_i$ is shown in the inset of Fig. 1(e). In the case of a strong oscillator (e.g. $\omega_0 = 1363$ cm$^{-1}$, see below), $\phi_{Im}(\omega_0)$ and $\theta_i$ are similar, but they may differ significantly for a weak mode.

The reflectance for $E \parallel b$ ($R_b$) and $E \parallel ac$ ($R_{ac}$) can be obtained by using the Fresnel equations for normal incidence [9]:

$$R_b = \left| (1 - \sqrt{\varepsilon_b}) \cdot (1 + \sqrt{\varepsilon_b})^{-1} \right|^2$$

$$R_{ac}(\chi) = \left| ((\hat{1} - \sqrt{\hat{\varepsilon}_{ac}}) \cdot (\hat{1} + \sqrt{\hat{\varepsilon}_{ac}})^{-1}) \begin{pmatrix} \cos\chi \\ \sin\chi \end{pmatrix} \right|^2 \quad (4)$$

In Eq. (4), $\chi$ denotes the angle between the polarization direction and the $x$ axis as shown in Fig. 1(a). In order to take the square root of a tensor, $\hat{\varepsilon}_{ac}$ first has to be rotated to a diagonal form, then the square root has to be taken for each element, and finally the resulting tensor is rotated back to its original basis.

The scalar $\varepsilon_b$ is determined by fitting the single spectrum $R_b$, while $\hat{\varepsilon}_{ac}$ is determined by fitting $R_{ac}(\chi)$ for three different values of $\chi$ simultaneously (we have measured $R_{ac}(\chi)$ for 14 different $\chi$ values). The fits yield an excellent description of the measured reflectivity. The fit parameters are listed in Tab. 1, and Fig. 1(e) shows Im$\{\varepsilon_{yy}\}$ and Im$\{\varepsilon_{xx}+\varepsilon_{zz}\}/2$, which is independent of $\theta$.

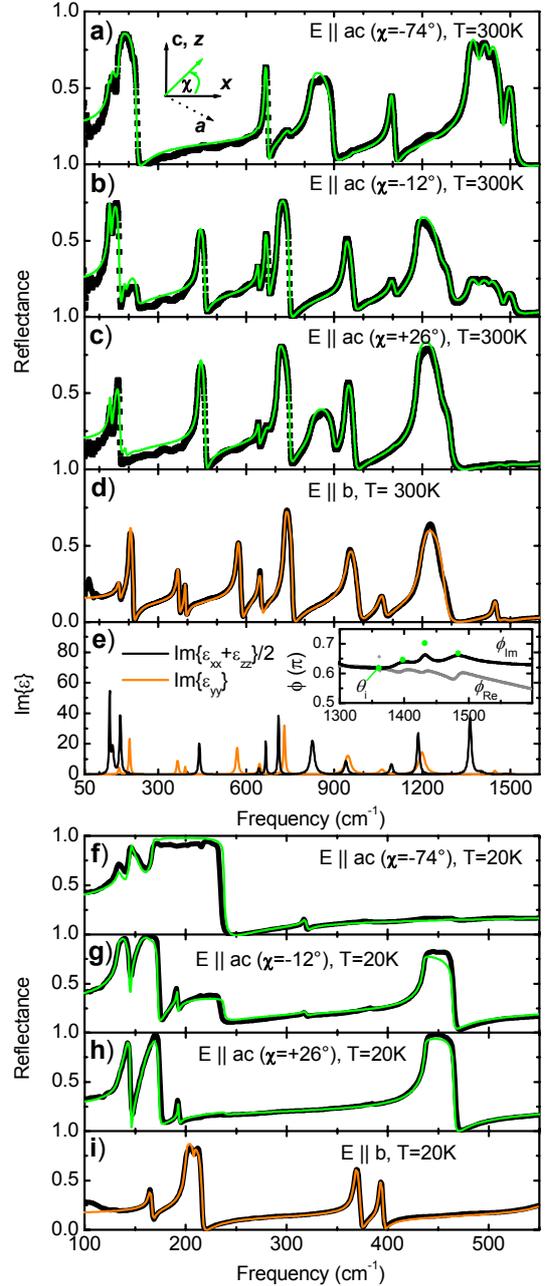

**Figure 1 a) – c)** Reflectivity spectra of BiB$_3$O$_6$ at $T = 300$ K for $E \parallel ac$ plane at different polarization angles $\chi$ (black) and fits using Eq. (4) (grey). **d)** Reflectivity at $T = 300$ K for E $\parallel b$ (black), fit using Eq. (4). **e)** Components of Im$\{\hat{\varepsilon}\}$ for the fitted parameters; comparison of $\theta_i$, $\phi_{Im}$, and $\phi_{Re}$ (inset). **f-h)** Reflectivity spectra of BiB$_3$O$_6$ at $T = 20$ K for $E \parallel ac$ plane at different polarization angles $\chi$ (black) and fits using Eq. (4) (grey). **i)** Reflectivity at $T = 20$ K for $E \parallel b$ (black), fit using Eq. (4).

**Table 1** Fit results for the reflectivity of BiB$_3$O$_6$ at $T = 300$ K using Eq. (4). Here, $\omega_0$ is the transverse frequency, $\omega_p$ the plasma frequency, $\gamma$ the damping, $\theta$ the angle between the dipole moment and the ***x*** axis (in case of the *B* modes), and $S = \omega_p^2/\omega_0^2$ denotes the oscillator strength. The high-frequency dielectric constants at $T = 300$ K are: $\varepsilon_{xx}^\infty = 3.5$, $\varepsilon_{yy}^\infty = 2.9$, $\varepsilon_{zz}^\infty = 3.3$ and $\varepsilon_{xz}^\infty = -0.3$. The four weak *B* modes at 1268, 1400, 1434, and 1486 cm$^{-1}$ probably correspond to multi-phonon excitations.

| *B* modes | | | | | *A* modes | | | |
|---|---|---|---|---|---|---|---|---|
| $\omega_0$ (cm$^{-1}$) | $\omega_p$ (cm$^{-1}$) | $\gamma$ (cm$^{-1}$) | $\theta$ (°) | S | $\omega_0$ (cm$^{-1}$) | $\omega_p$ (cm$^{-1}$) | $\gamma$ (cm$^{-1}$) | S |
| 136 | 236 | 3.9 | 151 | 3.021 | 168 | 71 | 7.0 | 0.179 |
| 146 | 215 | 9.4 | 132 | 2.185 | 202 | 164 | 5.5 | 0.659 |
| 172 | 291 | 6.5 | 89 | 2.862 | 367 | 151 | 7.0 | 0.169 |
| 190 | 50 | 8.3 | 33 | 0.069 | 393 | 93 | 4.4 | 0.056 |
| 441 | 333 | 6.0 | 18 | 0.570 | 570 | 305 | 9.2 | 0.286 |
| 644 | 173 | 5.8 | 175 | 0.073 | 646 | 208 | 9.4 | 0.104 |
| 667 | 328 | 3.7 | 131 | 0.241 | 731 | 396 | 6.6 | 0.293 |
| 711 | 514 | 4.8 | 15 | 0.523 | 947 | 522 | 23.0 | 0.304 |
| 826 | 753 | 15.6 | 77 | 0.831 | 1062 | 256 | 20.0 | 0.058 |
| 940 | 448 | 12.1 | 6 | 0.227 | 1199 | 668 | 25.6 | 0.310 |
| 1095 | 415 | 11.5 | 111 | 0.143 | 1448 | 208 | 12.3 | 0.021 |
| 1187 | 792 | 9.7 | 16 | 0.445 | 1484 | 129 | 30.3 | 0.008 |
| 1363 | 1028 | 10.5 | 112 | 0.568 | | | | |
| 1268 | 156 | 49.7 | 19 | 0.015 | | | | |
| 1400 | 314 | 21.4 | 117 | 0.050 | | | | |
| 1434 | 101 | 12.9 | 127 | 0.005 | | | | |
| 1486 | 169 | 19.4 | 121 | 0.013 | | | | |

In case of ***E*** ∥ ***b***, where 13 *A* modes are expected, we find 11 strong IR modes and a series of weaker features. Most of the latter can be interpreted as multi-phonon excitations, e.g., a very weak feature at 272 cm$^{-1}$, which can be attributed to an overtone of the B mode at 136 cm$^{-1}$ (note that $B \otimes B$ yields *A* symmetry). In order to determine which of the weak features corresponds to a fundamental phonon mode, it is helpful to compare our data with Raman results [6,7]. In a recent room-temperature Raman study [6], 12 *A* modes have been observed. Ten of these modes are found both in the Raman and in the IR data (the transverse frequencies agree within a few wave numbers). Combined with the two additional Raman modes at 1294 and 1488 cm$^{-1}$ and with the strongly IR-active mode at 1062 cm$^{-1}$, we end up with 13 *A* modes, as expected. The Raman peak at 1488 cm$^{-1}$ corresponds to a weak IR feature (at 1484 cm$^{-1}$). However, the Raman mode at 1294 cm$^{-1}$ [6] was neither observed in our IR data nor in the Raman data of Ref. [7]. Other features, which are weak in both spectroscopies, can be attributed to multi-phonon excitations (e.g. the peak at 272 cm$^{-1}$ discussed above) or to a polarizer leakage (e.g. the Raman mode at 443 cm$^{-1}$ [6] corresponds to a strongly IR-active *B* mode).

In case of the ***ac*** plane, each phonon mode shows a different orientation $\theta$ of the dipole moment. This produces complex patterns in $R_{ac}(\chi)$ which can not be described by a simple Drude-Lorentz model. For the description of $R_{ac}(\chi)$ we used 17 oscillators (see Tab. 1), in contrast to the 14 *B* modes predicted. The use of four oscillators below 250 cm$^{-1}$ is based on the 20 K data (see Fig. 1(f-h)). At high frequencies, the four modes at 1268, 1400, 1434, and 1486 cm$^{-1}$ are very weak and show rather large values of the damping $\gamma$. These four modes probably have to be attributed to multi-phonon excitations. Note that weak features may have a significant influence on the reflectivity if they are located on top of a *Reststrahlenband*. In case of the *B* modes the discrepancies between IR and Raman data [6,7] are much larger than for the *A* modes because transverse and longitudinal modes mix for ***E*** ∥ ***ac***. We have been able to determine the transverse eigenfrequencies and the orientation $\theta$ of the dipole moments by using the generalized Drude-Lorentz model (Eq. 3). In order to determine all 14 *B* modes and to distinguish two-phonon excitations from the one-phonon modes, detailed lattice-dynamical calculations are required.